\documentstyle[aps]{revtex}
\begin{document}
\draft
\title{Finite temperature Drude weight of the \\
one dimensional spin 1/2 Heisenberg model}
\author{X. Zotos}
\address{
Institut Romand de Recherche Num\'erique en Physique des
Mat\'eriaux (IRRMA), \\
INR-Ecublens, CH-1015 Lausanne, Switzerland}
\date{received: 21st August 1998}
\bigskip\bigskip
\maketitle
\begin{abstract}
Using the Bethe ansatz method, the zero frequency contribution (Drude weight)
to the spin current correlations is analyzed for the easy
plane antiferromagnetic Heisenberg model.
The Drude weight is a monotonically decreasing function of temperature
for all $0\leq \Delta \leq 1$, it approaches the zero temperature value
with a power law and it appears to vanish for all finite
temperatures at the isotropic $\Delta=1$ point.
\end{abstract}


The low frequency dynamics in one dimensional spin chains is a long standing 
problem. It has recently attracted a renewed interest, 
partly due to the fabrication of 
excellent quasi-one dimensional, spin 1/2 magnetic materials as $Sr_2CuO_3$ and 
$CuGeO_3$. Detailed NMR experiments\cite{tag} revealed an unusually high 
value of the spin diffusion constant and nearly ballistic behavior.

The first issue on the question of spin diffusion is the zero frequency 
contribution (or Drude weight) to the dynamic spin current correlations 
at finite temperatures.  If the Drude weight turns 
out to be finite, then the current correlations do not decay to zero at long 
times, implying ideal conducting behavior.
If they decay to zero, the question still remains open whether they decay 
fast enough so that transport coefficients can be defined.
Several numerical studies have been devoted to the analysis of  
the diffusive behavior in the Heisenberg model\cite{zp,fm,nz,nma} with 
suggestive but not conclusive results.

In relation to this problem, it has been proposed that the integrability of
the spin 1/2 Heisenberg model implies pathological spin dynamics and
presumably the absence of spin diffusion\cite{zp,mcc2}. 
A straightforward demonstration on the way in which
conservation laws, characterizing integrable systems, might affect the 
long time dynamics was pointed out in reference\cite{znp}. 
There it was shown that in several quantum
integrable models the uniform current correlations do not decay to zero
at long times due to the overlap of the currents to conserved 
quantities. Unfortunately, this simple idea turned out to be
insufficient for deciding about the decay of spin currents
in the spin 1/2 Heisenberg model at zero magnetic field.

On the other hand, a new method was proposed recently by Fujimoto and 
Kawakami\cite{fk} that allows the direct 
analytical evaluation of the Drude weight at finite temperatures.
This procedure is based on the calculation of finite size 
corrections of the energy eigenvalues obtained by the Bethe ansatz 
method \cite{bm}.
The analysis starts from a convenient expression for the 
finite temperature Drude weight 
as the thermal average of curvatures of energy levels in a Hamiltonian
subject to a fictitious flux coupled to the hopping or spin 
flipping term\cite{kohn,czp}. 
Note that the anisotropic Heisenberg model is equivalent to 
the model of spinless fermions interacting with nearest neighbor interaction 
using the Jordan-Wigner transformation\cite{emery}. The direct analogy 
between charge and spin current correlations, suggests also the use 
of the name ``Drude weight'' in the context of spin correlations.

In this work, we calculate the Drude weight for the antiferromagnetic 
Heisenberg model using the procedure proposed in references\cite{bm,fk}.
The formulation and notation of the thermodynamic Bethe ansatz equations 
by Takahashi and Suzuki\cite{ts} will be closely followed.
This construction is based on the string assumption for the excitations 
and it is particularly complex for arbitrary values of the anisotropy 
parameter $\Delta$. The allowed type of strings are constrained by the 
normalizability of the wavefunctions\cite{str}. 
Therefore, for simplicity and without loss of generality, the analysis 
will be limited 
at $\Delta=\cos(\pi/\nu)$, $\nu$=integer, where only a finite number of
string excitations is allowed.

The results presented here are in good agreement with numerical results 
obtained by exact diagonalization of the Hamiltonian matrix on finite size 
lattices\cite{nz}. They lend support both to
the string construction and the novel procedure for calculating the Drude
weight from finite size corrections to the Bethe ansatz eigenvalues.

\bigskip
The XXZ anisotropic Heisenberg Hamiltonian for a chain of $N$ sites
with periodic boundary conditions $S_{N+1}^a=S_1^a$ is given by:

\begin{equation}
H=J \sum_{i=1}^N(S_i^x S_{i+1}^x+S_i^y S_{i+1}^y +\Delta S_i^z S_{i+1}^z)
\label{heis}
\end{equation}
where $S_i^a=\frac{1}{2}\sigma_i^a$,
$\sigma_i^a$ are the Pauli spin operators with components
$a=x,y,z$ at site $i$.
The region $0\leq \Delta \leq 1$ is parametrized by $\Delta=\cos\theta$, 
$\theta=\pi/\nu$,
$\nu=\rm integer$. The pseudomomenta $k_\alpha$ and phase shifts
$\phi_{\alpha\beta}$ characterizing the Bethe ansatz
wavefunctions are expressed in terms of the rapidities $x_\alpha$:

\begin{eqnarray}
\cot(\frac{k_{\alpha}}{2})=\cot(\frac{\theta}{2})
\tanh(\frac{\theta x_{\alpha}}{2}), \nonumber\\  
\cot(\frac{\phi_{\alpha\beta}}{2})=\cot(\frac{\theta}{2})
\tanh(\frac{\theta (x_{\alpha}-x_{\beta})}{2}).
\label{param}
\end{eqnarray}

For M down spins and N-M up spins the energy E and momentum K are given by:
\begin{equation}
E=J\sum_{\alpha=1}^M (\cos k_{\alpha}-\Delta),~~~~
K=\sum_{\alpha=1}^M k_{\alpha}.
\label{enem}
\end{equation}

Coupling the spin flipping term to a fictitious flux $\phi$, the Hamiltonian
becomes:
\begin{equation}
H=J \sum_{i=1}^N (\frac{1}{2}e^{i\phi}\sigma_i^+ \sigma_{i+1}^-+ h.c.)
+\Delta S_i^z S_{i+1}^z).
\label{hphi}
\end{equation}
The finite temperature Drude weight $D$ can then be calculated by\cite{czp}:

\begin{equation}
D =\frac{1}{N} \sum_n p_n
\frac{1}{2} \frac {\partial ^2 E_n (\phi) }{\partial \phi^2}|_{\phi
\rightarrow 0} .
\label{chst}
\end{equation}
where $E_n$ are the eigenvalues of the Hamiltonian and $p_n$ the
corresponding Boltzmann weights.
Imposing periodic boundary conditions on the Bethe ansatz wavefunctions
the following relations are obtained:
\begin{equation}
\bigl\lbrace \frac{\sinh \frac{1}{2}\theta(x_\alpha+i)}
{\sinh\frac{1}{2}\theta(x_\alpha-i)}\bigr\rbrace^N
=-e^{i\phi N}\prod_{\beta=1}^M
\bigl\lbrace \frac{\sinh \frac{1}{2}\theta(x_\alpha-x_\beta+2i)}
{\sinh\frac{1}{2}\theta(x_\alpha-x_\beta-2i)}\bigr\rbrace;
~~~\alpha=1,2,...M.
\label{pbce}
\end{equation}

In the thermodynamic limit, the solutions of equations (\ref{pbce}) are 
grouped into
strings of order $n_j, j=1,...,\nu$ and parity $v_j=+~ {\rm or}~ -$. 
For $\theta=\pi/\nu$ the allowed
strings are of order $n_j=j, j=1,...,\nu-1$ and parity $v_j=+$ of the form:
\begin{equation}
x_{\alpha,+}^{n,k}=x_{\alpha}^n+(n+1-2k)i+O(e^{-\delta N});~~~k=1,2,...n,
\label{s1}
\end{equation}
and strings of order $n_{\nu}=1$ and parity $v_{\nu}=-$ of the form,
\begin{equation}
x_{\alpha,-}=x_{\alpha}+i\nu+O(e^{-\delta N}),~~~\delta > 0. 
\label{s2}
\end{equation}

Multiplying the terms in equation (\ref{pbce}) corresponding to 
different members of a string and taking the logarithm we obtain:
\begin{equation}
Nt_j(x_{\alpha}^j)=2\pi I_{\alpha}^j+\sum_{k=1}^{\infty}\sum_{\beta=1}^{M_k}
\Theta_{jk}(x_{\alpha}^j-x_{\beta}^k)+
n_j\phi N;~~~ \alpha=1,2,...M_j,
\label{basic}
\end{equation}
$I_{\alpha}^j$ are integers (or half-integers) and $M_k$ is the number 
of strings of type $k$,
\begin{eqnarray}
t_j(x)&=&f(x;n_j,v_j), \nonumber\\
\Theta_{jk}(x)&=&f(x;|n_j-n_k|,v_jv_k)+
f(x;n_j+n_k,v_jv_k)+ \nonumber\\
&2&\sum_{i=1}^{Min(n_j,n_k)-1}f(x;|n_j-n_k|+2i,v_jv_k), \nonumber\\
f(x;n,v)&=&2v\tan^{-1}\lbrack(\cot(n\pi/2\nu))^v\tanh(\pi x/2\nu)\rbrack.
\nonumber
\end{eqnarray}

Following reference\cite{bm} the finite size corrections to the 
energy eigenvalues for a system of size $N$ are calculated by introducing 
the function $g_{1j},g_{2j}$:
\begin{equation}
x_N^j=x_{\infty}^j+\frac{g_{1j}}{N}+\frac{g_{2j}}{N^2}.
\label{g}
\end{equation}
where $x_N^j(x_{\infty}^j)$ are the rapidities for a system of 
size $N(\infty)$.
Next, we expand equations (\ref{basic}) to orders of $1/N$ and in the 
thermodynamic limit introduce the densities of excitations $\rho_j$ and 
hole densities $\rho_j^h$. The sums over the pseudomomenta are replaced by 
integrals over excitation densities plus boundary terms using the 
Euler-Maclaurin formula.

To $O(1)$ we recover the integral equations for the excitation densities 
in the thermodynamic limit\cite{ts}:
\begin{equation}
a_j=\lambda_j(\rho_j+\rho_j^h)+\sum_k T_{jk}\ast \rho_k.
\label{aj}
\end{equation}
$\ast$ denotes the convolution  
$a\ast b(x)=\int_{-\infty}^{+\infty} a(x-y)b(y)dy$, 
$T_{jk}(x)=(1/2\pi)d\Theta_{jk}(x)/dx$ and $a_j(x)=(1/2\pi)dt_j(x)/dx$.
The sum over $k$ is constrained to the allowed strings, given in our case by  
the equations (\ref{s1},\ref{s2}) and $\lambda_j=1, j=1,...,\nu-1$,  
$\lambda_{\nu}=-1$

\bigskip
\noindent
To $O(1/N)$:
\begin{equation}
\lambda_j g_{1j}(\rho_j+\rho_j^h)=-\sum_k T_{jk}\ast (g_{1k}\rho_k)+
\frac{n_j\phi}{2\pi},
\label{g1j}
\end{equation}

\bigskip
\noindent
To $O(1/N^2)$
\begin{eqnarray}
\lambda_j g_{2j}(\rho_j+\rho_j^h)+\sum_k T_{jk}\ast (g_{2k}\rho_k)=\nonumber\\
\frac{1}{2}\frac{d}{dx}\lbrace \lambda_j g_{1j}^2(\rho_j+\rho_j^h)+
\sum_k T_{jk}\ast (g_{1k}^2\rho_k)\rbrace +\nonumber\\
{\rm boundary~~terms}
\label{g2j}
\end{eqnarray}

Minimizing the free energy we obtain the standard Bethe ansatz equations for 
the equilibrium densities $\eta_j=\rho_j^h/\rho_j$ at temperature 
$T(\beta=1/\kappa_B T)$:
\begin{equation}
\ln \eta_j=-2\nu\sin(\pi/\nu)Ja_j\beta+\sum_k \lambda_kT_{jk}\ast 
\ln(1+\eta_k^{-1})
\label{eqe}
\end{equation}
These relations define the temperature dependent effective dispersions 
$\epsilon_j=(1/\beta)\ln(\rho_j^h/\rho_j)$.
In the string representation the energy is given by:
\begin{equation}
E=N\sum_{j=1}^{\infty}\int_{-\infty}^{+\infty}dx
(-2\nu\sin(\frac{\pi}{\nu})Ja_j(x))\rho_j(x)
\label{ene}
\end{equation}
Expanding this expression for the energy 
we find that the first order correction in $1/N$ vanishes. So, the 
second derivative with respect to $\phi$ of the second order correction 
gives us the final expression for the Drude weight:
\begin{equation}
D=\frac{1}{2}\sum_j\int_{-\infty}^{+\infty}dx 
\bigl\lbrack (\rho_j+\rho_j^h)
\frac{\partial g_{1j}}{\partial \phi}\bigr\rbrack^2 
\frac{d}{dx} (\frac{-1}{1+e^{\beta\epsilon_j}})
(\frac{1}{\rho_j+\rho_j^h}
\frac{d\epsilon_j}{dx})
\label{d}
\end{equation}
This expression is formally similar to the one obtained in reference \cite{fk} 
for the Drude weight in the Hubbard model.
It has an elegant interpretation by comparing it to the analogous  
expression for independent fermions.
Taking the second derivative of the free energy with respect to 
the flux $\phi$ we find:
\begin{equation}
\frac{\partial^2 F}{\partial \phi^2}=
\sum_{\mu} <n_{\mu}>\frac{\partial^2 \epsilon_{\mu}}{\partial\phi^2}
-\beta\sum_{\mu} <n_{\mu}>(1-<n_{\mu}>) 
(\frac{\partial \epsilon_{\mu}}{\partial\phi})^2
\label{ff}
\end{equation}
where $<n_{\mu}>$ is the Fermi-Dirac distribution for particles with 
dispersion $\epsilon_{\mu}$.
Considering that the left hand side (the persistent current 
susceptibility ) vanishes in the thermodynamic 
limit for any finite temperature and that the first term in the right 
hand side is equal to $2ND$, we find that:
\begin{equation}
D\simeq
\frac{\beta}{2N}\sum_{\mu} <n_{\mu}>(1-<n_{\mu}>) 
(\frac{\partial \epsilon_{\mu}}{\partial\phi})^2|_{\phi\rightarrow 0}
\label{df}
\end{equation}
Rewriting equation (\ref{d}) we arrive at a similar expression:
\begin{equation}
D=\frac{1}{2}\beta\sum_j\int_{-\infty}^{+\infty}dx (\rho_j+\rho_j^h)
<n_j>(1-<n_j>) (\frac{\partial \epsilon_j}{\partial x}
\frac{\partial g_{1j}}{\partial\phi})^2
\label{dh}
\end{equation}
with $<n_j>=1/(1+e^{\beta\epsilon_j})$.
So the Bethe ansatz expression for the Drude weight resembles that of 
independent fermion-like excitations. 

To obtain the distributions $\rho_j,\rho_j^h$ and 
$\frac{\partial g_{1j}}{\partial \phi}$, the coupled integral 
equations (\ref{eqe}),(\ref{aj}),(\ref{g1j}) are numerically solved by 
iteration.

In Fig.~1, $D$ is shown as a function of $\Delta$ for $2\leq \nu\leq 16$ 
($0\leq\Delta < 0.98$) and different characteristic temperatures. 
The main result is that the Drude weight $D$ is a monotonically decreasing 
function of $\Delta$ and temperature. At $T=0$, 
$D=\frac{\pi}{8}\frac{\sin(\pi/\nu)}
{\frac{\pi}{\nu}(\pi-\frac{\pi}{\nu})}$
\cite{ss}.
Most interestingly, $D$ seems to vanish at all temperatures  
for $\Delta=1$. This result excludes an ideal conducting behavior 
for the isotropic Heisenberg model. 
Still, an anomalously slow long time decay of the current correlation functions 
could lead to pathological low frequency dynamics and non-diffusive behavior.
Furthermore, the vanishing of the Drude weight at the isotropic point suggests 
that it remains zero at all temperatures in the region $\Delta >1$, 
the easy axis case (or insulating state in the fermionic model). This 
conclusion is in accord with the numerical results of reference \cite{zp}.
We should note that the numerical investigation close to the isotropic 
point is somewhat difficult as the number of equations to solve diverges.

In the high temperature limit ($\beta\rightarrow 0$), $D$ is proportional 
to $\beta$. The constant of 
proportionality $C_{jj}$, equal to the long time asymptotic value of the 
current correlations\cite{znp}, is compared with 
results obtained in reference\cite{nz} by exact 
diagonalization of the Hamiltonian on finite size 
lattices extrapolated to the infinite size limit.
The quantitative agreement obtained lends support to the assumptions 
involved in the whole Bethe ansatz procedure for calculating thermodynamic 
properties and finite size corrections.

The next observation is that the Drude weight approaches the zero 
temperature value with a power law of the form:
\begin{equation}
D(T)=D(T=0)-{\rm const.}T^{\alpha},~~~\alpha=\frac{2}{\nu-1}
\label{pow}
\end{equation}
To indicate this point, in Fig.~2, $D(T=0)-D(T)$ is shown for $\nu=3,...,6$ 
in a logarithmic plot along with lines of slope $\alpha$.
Note that the exponent $\alpha$ is half that 
for the low temperature spin susceptibility as obtained by 
Abelian bosonization\cite{eat}. 
It is also consistent with the value $\alpha=2$ for free fermions ($\nu=2$).

\bigskip
The results presented above, concern only the zero frequency contribution  
to the spin current correlations. A reliable method for studying the 
{\it low frequency} behavior in integrable quantum many body systems 
(and the influence of non-integrable perturbations) remains a challenging 
problem. 

\acknowledgments
We would like to thank F. Naef, P. Prelov\v sek and M. Long for useful 
discussions.
This work was supported by the Swiss National Science Foundation grant
No. 20-49486.96, the University of Fribourg, the University of Neuch\^atel 
and the $\Pi$ENE$\Delta$ 95 research program.

\begin{figure}
\caption{
$D(\Delta)$ evaluated at the points $\nu=3,...,16$ and various 
temperatures. The continuous line 
is the high temperature proportionality constant $C_{jj}=D/\beta$.
The $\diamond$ indicate exact diagonalization results 
from reference (4).}
\label{one}
\caption{
$D(T=0)-D(T)$ at different temperatures ($\diamond$'s) in a 
logarithmic scale.  The lines indicate slopes $\alpha=2/(\nu-1)$.}
\label{two}

\end{figure}

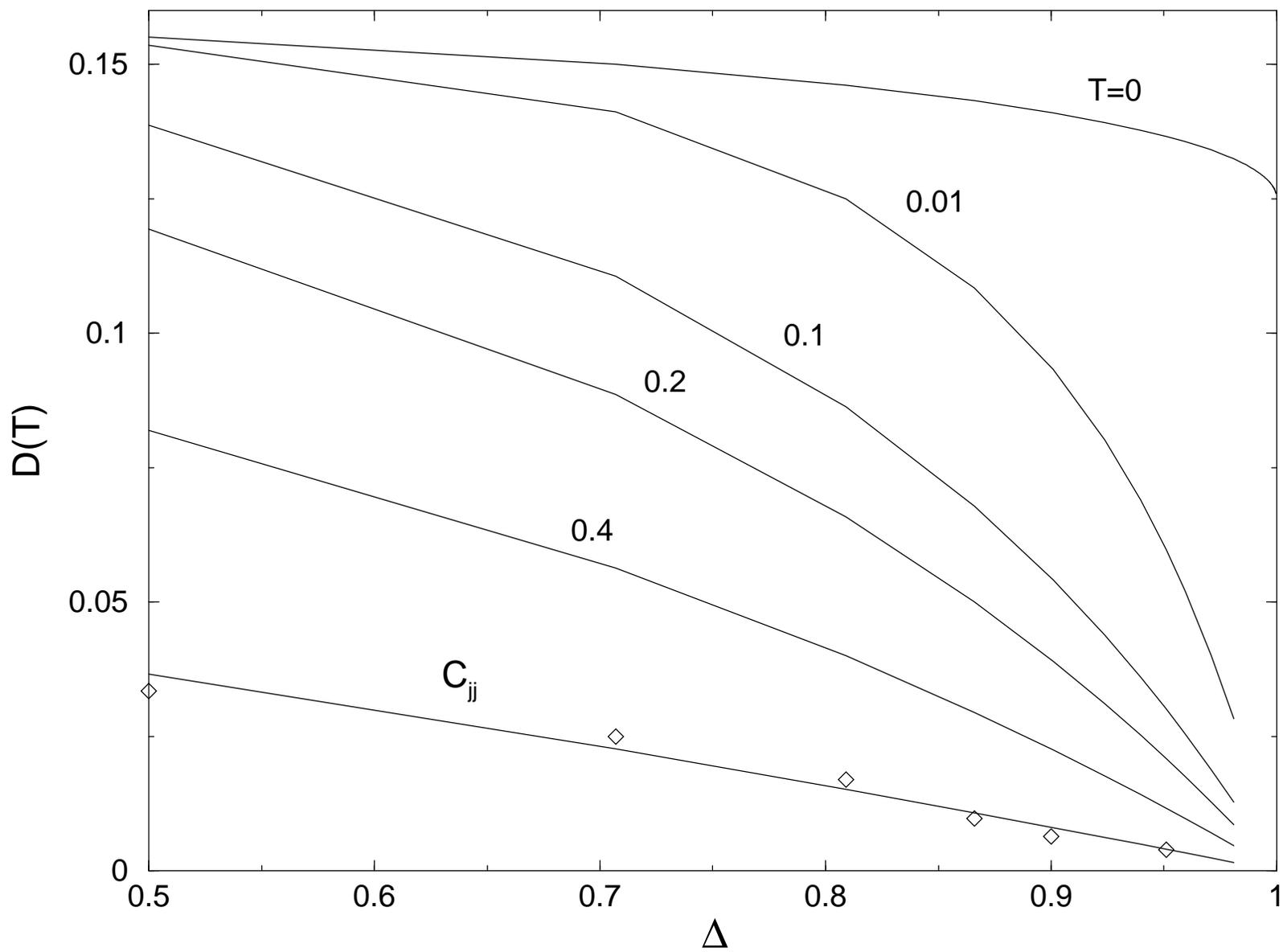
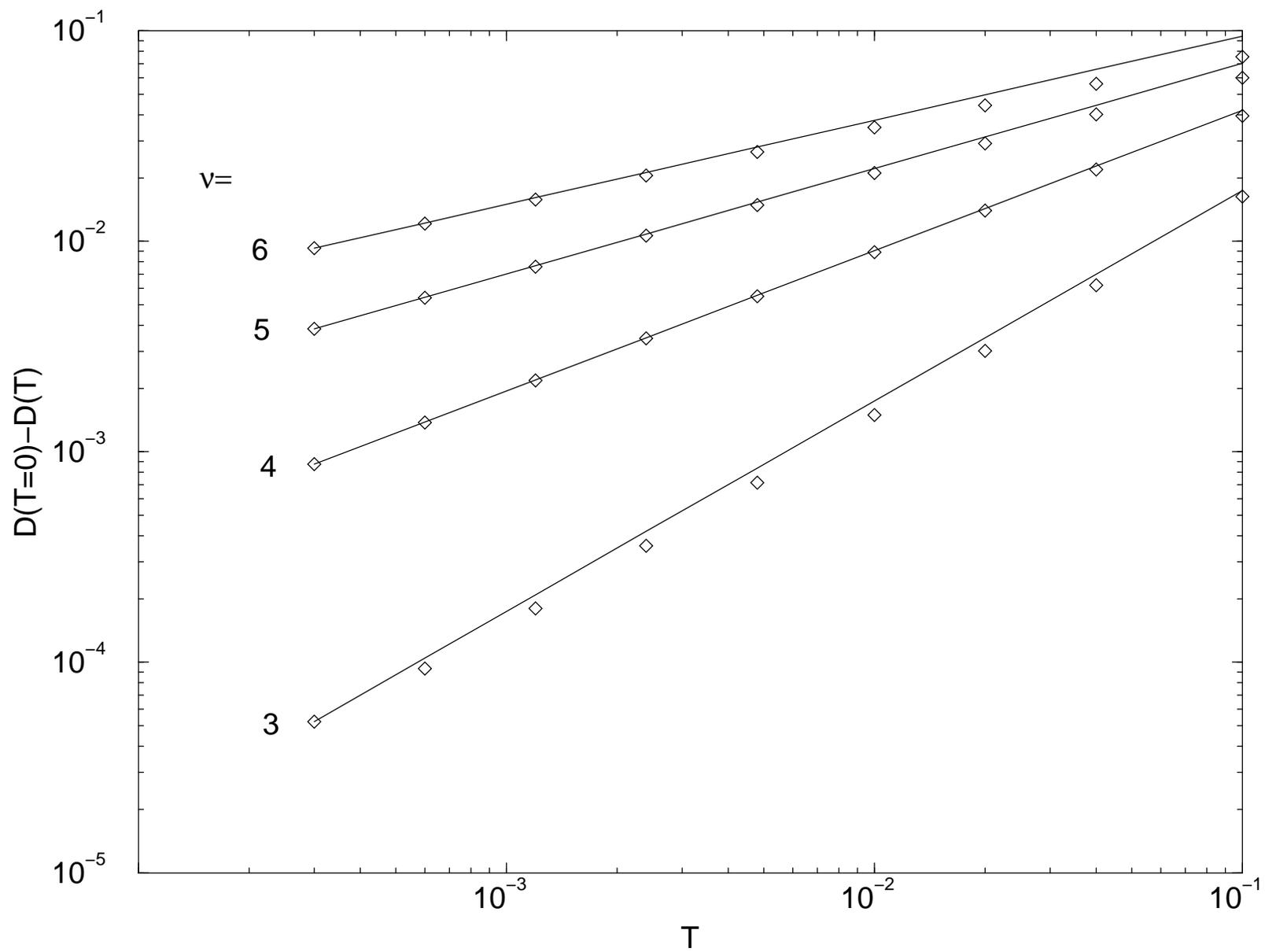
\end{document}